\documentclass[11pt,draftcls,onecolumn]{IEEEtran}

\usepackage{amsmath}
\usepackage{amssymb}
\usepackage{amsthm}
\usepackage{graphicx}
\usepackage{cases}
\usepackage{subfig}
\usepackage{enumerate}
\usepackage{algorithmic}
\usepackage{algorithm}
\usepackage{datetime}
\usepackage{booktabs}
\usepackage{multirow}
\usepackage{acronym}
\usepackage{ifthen}

\newboolean{Draft}
\setboolean{Draft}{true}

\acrodef{CS}{Compressive Sensing}
\acrodef{Rell1}[R$\ell_1$CC]{Reweighted $\ell_1$ with Clipping Constraints}
\acrodef{TP}{Trivial Pursuit}
\acrodef{TPCC}{Trivial Pursuit with Clipping Constraints}
\acrodef{DFT}{Discrete Fourier Transform}
\acrodef{OMP}{Orthogonal Matching Pursuit}
\acrodef{DCT}{Direct Cosine Transform}

\usepackage{xspace}
\makeatletter
\DeclareRobustCommand\onedot{\futurelet\@let@token\@onedot}
\def\@onedot{\ifx\@let@token.\else.\null\fi\xspace}
 
\def\ie{{i.e}\onedot}

\def\etal{{et al}\onedot}
\makeatother

\DeclareMathOperator*{\argmin}{argmin}
\DeclareMathOperator*{\argmax}{argmax}
\DeclareMathOperator*{\supp}{supp}

\DeclareMathOperator{\DFT}{DFT}

\newcommand{\real}{\mathbb{R}}
\newcommand{\complex}{\mathbb{C}}

\ifthenelse{\boolean{Draft}}{
\captionsetup{font={small,singlespacing}}
}{}

\author{Alejandro J. Weinstein and Michael B. Wakin\thanks{Department of Electrical Engineering and Computer Science, Colorado School of Mines, Golden, CO 80401. Email: \{aweinste,mwakin\}@mines.edu. This work was partially supported by AFOSR grant FA9550-09-1-0465.

AJW and MBW thank Mike Mooney, Norman Facas, and Nathan Toohey for inspiring us to look at this problem.
}
}

\title{Recovering a Clipped Signal in Sparseland}

\begin{document}
\maketitle

\ifthenelse {\boolean{Draft}}{\vspace{-10ex}}{}

\begin{abstract}
In many data acquisition systems it is common to observe signals whose amplitudes have been clipped.
We present two new algorithms for recovering a clipped signal by leveraging the model assumption that the underlying signal is sparse in the frequency domain.
Both algorithms employ ideas commonly used in the field of Compressive Sensing; the first is a modified version of Reweighted $\ell_1$ minimization, and the second is a modification of a simple greedy algorithm known as Trivial Pursuit.
An empirical investigation shows that both approaches can recover signals with significant levels of clipping.
\end{abstract}


\section{Introduction}

In many practical situations, either because a sensor has the wrong dynamic range or because signals arrive that are larger than anticipated, it is common to record signals whose amplitudes have been clipped.
Any method for restoring the values of the clipped samples must---implicitly or explicitly---assume some model for the structure of the underlying signal.
For example, one of the first attempts to ``de-clip'' a signal was the work of Abel and Smith \cite{Abel1991}, who assumed that the underlying signal had limited bandwidth relative to the sampling rate (i.e., that it was oversampled) and recovered the original signal by solving a convex feasibility problem. 
Godsill \etal~\cite{Godsill2001} later tackled the de-clipping problem using a parametric model and a Bayesian inference approach.
Along the same lines, Olofsson \cite{Olofsson2005} proposed a maximum \emph{a posteriori} estimation technique for restoring clipped ultrasonic signals based on a signal generation model and a bandlimited assumption.

Meanwhile, recent research in fields such as \ac{CS}~\cite{Candes2006a} has shown the incredible power of sparse models for recovering certain signal information.
Many signals can be naturally assumed to be sparse in that they have few non-zero coefficients when expanded in a suitable basis; the name ``Sparseland'' has been informally used to describe the broad universe of such signals~\cite{Elad2010a}.
Although a typical CS problem involves an incomplete set of random measurements (as opposed to a complete---but clipped---set of deterministic samples), sparse models have made a limited appearance in the de-clipping literature.
In particular, Gemmeke \cite{Gemmeke2010} \etal imputed noisy speech features by considering the spectrogram of the signal as an image with missing samples, represented the spectrogram in terms of an overcomplete dictionary, and used sparse recovery techniques to recover the missing samples.
Using the model assumption that the underlying signal is sparse in 
an overcomplete harmonic dictionary, Adler \etal \cite{Adler2011} later adapted the \ac{OMP} \cite{Elad2010a} recovery algorithm from CS into a de-clipping algorithm that they call \emph{constraint}-\ac{OMP}.

In this paper, we present two methods for de-clipping a signal under the assumption that the original signal is sparse in the frequency domain, i.e., that it can be represented as a concise sum of harmonic sinusoids.
This model is general enough to embrace a wide set of signals that could be recorded from certain communication systems, resonant physical systems, etc.
This model is also commonplace in the CS literature, particularly in settings where random measurements are collected in the time domain.

Although the measurements we consider are not randomly collected,\footnote{In fact they are ``adversarial'', in that clipping eliminates the samples with the highest energy content.} we do find that certain ideas from the field of \ac{CS} can be leveraged.
In particular, we have modified several CS algorithms in an attempt to account for the clipping constraints.
Among the methods that we have tried, the two with the best performance are a modified version of Reweighted $\ell_1$ minimization
\cite{Candes2008b} and a modified version of the Thresholding algorithm \cite{Elad2010a}, also known as \ac{TP} \cite{Baron2005}.
This is surprising since \ac{TP}, a very simple greedy algorithm, is one of the poorest performing algorithms in conventional \ac{CS} problems~\cite{Elad2010a}.
We also show that, when tested on frequency sparse signals, these two methods outperform \emph{constraint}-\ac{OMP}.

\section{Preliminaries}\label{sec:prem}

\subsection{Problem definition}

Let $x \in \real^N$ be a $K$-sparse signal in the Fourier domain, \ie, $x=\Psi\alpha$ and $\|\alpha\|_0=K$, where $\Psi$ is the $N \times N$ inverse \ac{DFT} matrix and $\|\cdot\|_0$ denotes the number of non-zero entries of a vector. Because of the Hermitian symmetry property of real signals, the sparsity level $K$ is in general twice the number of harmonics in $x$.\footnote{The exceptions are harmonics of frequency 0 or $\pi$, which contribute only one \ac{DFT} coefficient each.} Let the clipped version of $x$ be $x_c$, where
\begin{equation*}
  x_c(n) = \begin{cases}
    C_u & \text{if $x(n) \geq C_u$}, \\
    C_l & \text{if $x(n) \leq C_l$}, \\
    x(n) & \text{otherwise},
  \end{cases}
\end{equation*}
and $C_u$ and $C_l$ are the known upper and lower clipping values, respectively. Our goal is to recover the original signal $x$ from the observed
clipped signal $x_c$.

Denote by $\Omega_u$ and $\Omega_l$ the index sets of the upper and lower clipped samples, respectively, and by $\Omega_{nc}$ the index set of the
non-clipped samples: $\Omega_u  = \{n | x_c(n)=C_u \}$, $\Omega_l  = \{n | x_c(n)=C_l \}$, and $\Omega_{nc} = (\Omega_u \cup \Omega_l)^c$.
Similarly, denote by $\Psi_u$ and $\Psi_l$ the matrices formed with the rows $i \in \Omega_u$ and $j \in \Omega_l$ of $\Psi$, respectively. We can write the non-clipped values of $x_c$ as $y = \Phi x$,
where $\Phi$ is a \emph{restriction operator} formed with the rows $j \in \Omega_{nc}$ of the $N \times N$ identity matrix.

\subsection{Basis Pursuit, Basis Pursuit with Clipping Constraints, and Reweighted $\ell_1$ with Clipping Constraints}

The canonical \ac{CS} method for recovering a sparse signal is known as Basis Pursuit~\cite{Elad2010a}. Given a set of non-clipped linear measurements $y = \Phi x = \Phi \Psi \alpha$, Basis Pursuit involves solving the following convex optimization problem:
\begin{equation}
  \label{eq:bp}
  \alpha = \argmin_{\alpha \in \complex^N}  \|W\alpha\|_1
  \quad\text{s.t.}\quad \Phi\Psi\alpha=y,
  \tag{BP}
\end{equation}
where $W$ is a diagonal weighting matrix with the norm of the columns of $\Phi\Psi$ in its main diagonal and zeros elsewhere. In the de-clipping problem, we also know that samples clipped by the upper limit must have values greater or equal than $C_u$, and samples clipped by the lower limit must have values smaller or equal than $C_l$. We can then propose a version of Basis Pursuit with clipping constraints:
\begin{equation}
  \label{eq:bpcc}
  \begin{split}
   &\alpha = \argmin_{\alpha \in \complex^N}  \|W\alpha\|_1   \\
    & \text{s.t.} \quad \Phi \Psi\alpha=y,\  \Psi_u\alpha \geq C_u,
    \ \Psi_l\alpha \leq C_l.  \\
\end{split} \tag{BPCC}
\end{equation}

Another technique commonly used in \ac{CS} is ``Reweighted $\ell_1$ minimization'' \cite{Candes2008b}. In its original formulation, this method iterates over a weighted version of \eqref{eq:bp}, adjusting the weights based on the solution obtained in the previous iteration. This method typically has better signal recovery performance than Basis Pursuit but at the expense of a higher computational load. We adapt this method to the de-clipping problem by replacing \eqref{eq:bp} at each iteration with \eqref{eq:bpcc}. We dub this method \ac{Rell1}. Algorithm \ref{alg:reweighted} shows the complete method.

\begin{algorithm}
\caption{Reweighted $\ell_1$ minimization with clipping constraints (\ac{Rell1})}
\label{alg:reweighted}
\fontsize{9}{9}\selectfont 
\begin{algorithmic}

\REQUIRE $\Phi$, $\Psi$, $\Psi_u$, $\Psi_l$, $y$, $C_l$, $C_u$, $\ell_{max}$, $\epsilon$,
$\delta$

\STATE $\ell= 1$, $W_i^{(1)} = 1$, $i=1,\ldots,N$
\REPEAT
\STATE $\alpha^{(\ell)} = \arg\min \|W^{(\ell)}\alpha\|_1$ \\ ~~~~~~~~ s.t. $\Phi\Psi\alpha=y$, $\Psi_u\alpha \geq C_u,$
  $\Psi_l\alpha \leq C_l$

\STATE $W_i^{(\ell+1)} = \frac{1}{|\alpha_i^{(\ell)}| + \epsilon}$, $i=1,\ldots,N$
\STATE $\ell= \ell+1$
\UNTIL{$\ell \geq \ell_{max} + 1$ \OR $\|\alpha^{(\ell)} - \alpha^{(\ell-1)} \|_2 <
  \delta$}
\ENSURE $\alpha^{\ell-1}$
\end{algorithmic}
\end{algorithm}

\begin{figure}[tb]
  \centering
  \subfloat[\label{fig:test1a}]{
    \includegraphics[width=0.5\columnwidth]{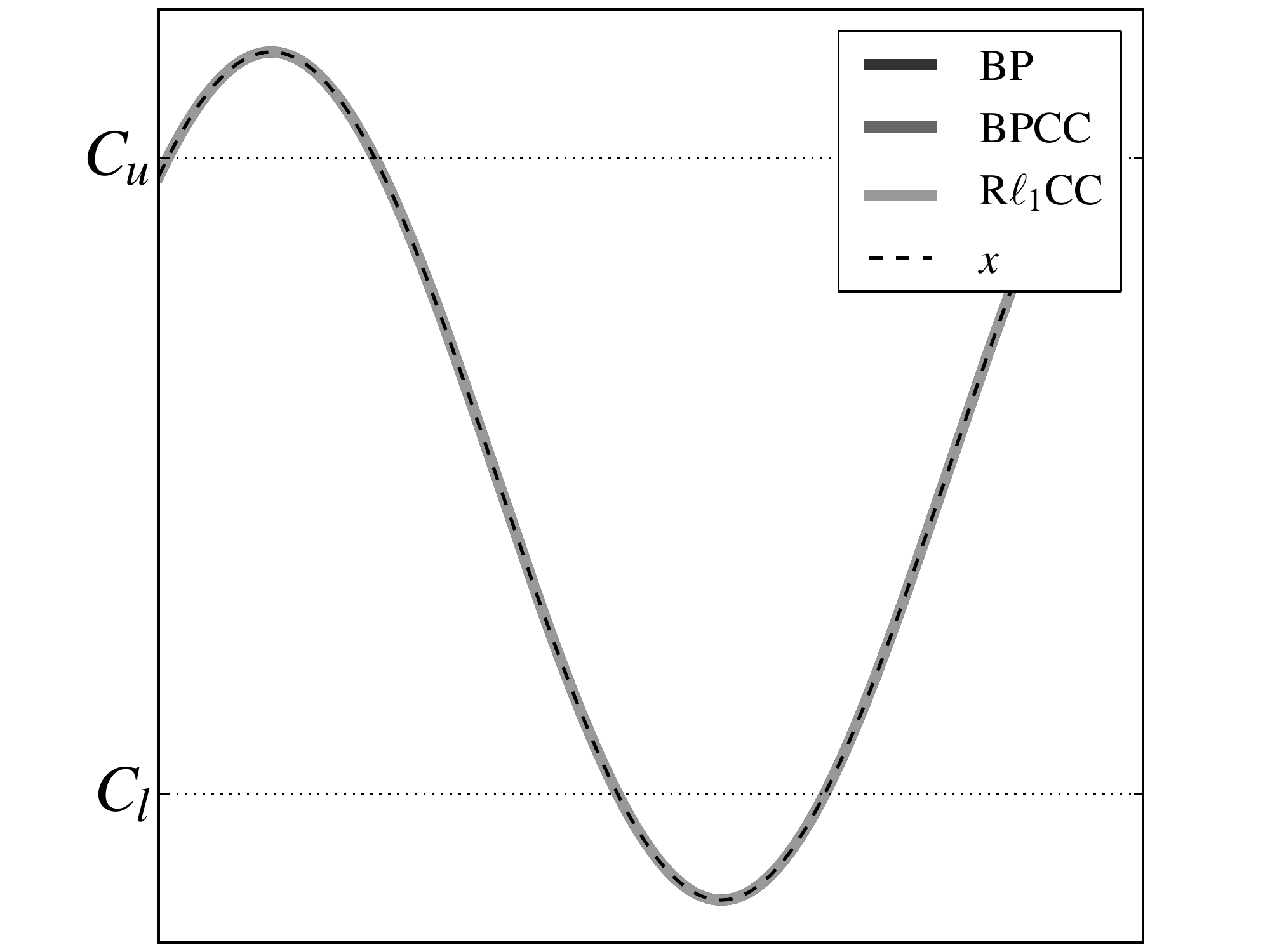}
  }
  \subfloat[\label{fig:test1b}]{
    \includegraphics[width=0.5\columnwidth]{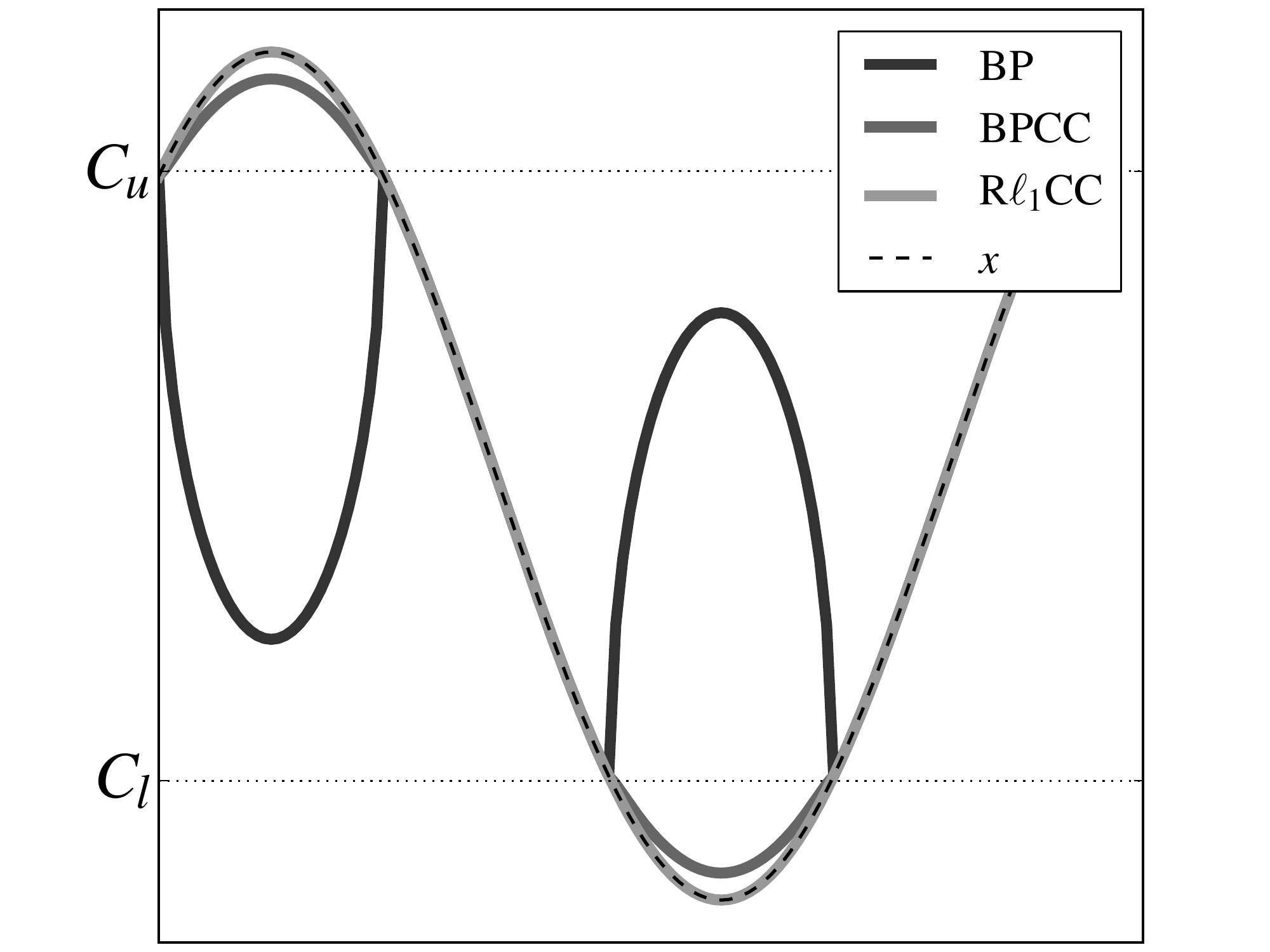}}
  \caption{\small\sl Reconstruction of $x(n) = \sin\left(2\pi n/N + \pi/4 \right)$ by
    \eqref{eq:bp}, \eqref{eq:bpcc}, and Algorithm 1 (\ac{Rell1}). (a)~Clipping level $\pm 0.75$. All three approaches recover the signal. (b)~Clipping level $\pm 0.72$. Only \ac{Rell1} recovers the signal exactly. \label{fig:sim_res} \vspace{-0.2in}}
\end{figure}

We test these three approaches with the signal $x(n) = \sin\left(2\pi n/N + \pi/4 \right)$ for $N=128$. Figure~\ref{fig:sim_res}\subref{fig:test1a} shows the result for a clipping level of $\pm 0.75$, at which there are 70 non-clipped samples, and Fig.~\ref{fig:sim_res}\subref{fig:test1b} shows the result for a clipping level of $\pm 0.72$, at which there are 66 non-clipped samples. These numbers of non-clipped samples\footnote{Due to the nature of this signal $x(n)$, it is not possible to set the clipping level so that the number of non-clipped samples is between 66 and 70.} correspond to the transition between the recovery and non-recovery zones of operation of \eqref{eq:bp} and \eqref{eq:bpcc}.

In this experiment and in others (see Sec.~\ref{sec:experiments}), we observe that adding clipping constraints to Basis Pursuit does not help to perfectly recover signals with lower clipping thresholds. \ac{Rell1}, on the other hand, can recover signals with more significant levels of clipping. This improvement of \ac{Rell1} over \eqref{eq:bp} and \eqref{eq:bpcc} is actually substantially better than is typically observed in \ac{CS}~\cite{Candes2008b}.

Thinking in terms of \ac{CS} principles, the Restricted Isometry Property (RIP)~\cite{Candes2006a} is commonly used for theoretical analysis of compressive measurement operators.
The RIP can be shown to hold with high probability for a randomly generated matrix with a small number of rows, and when it holds, such a matrix can be used to exactly recover any sparse signal (up to a certain sparsity level).
This perspective is not the right one to analyze the de-clipping problem, however, and it cannot be used to explain why \eqref{eq:bp} and \eqref{eq:bpcc} fail in the previous example.
First, since the matrix $\Phi\Psi$ is not random, we cannot use any of the standard probabilistic tools to predict whether it will satisfy the RIP.
Second, while the RIP guarantees that a fixed measurement matrix can be used to recover {\em any} sparse signal, in the de-clipping problem the matrix $\Phi\Psi$ is relevant {\em only} for the small set of signals that, when clipped, actually produce the samples given by this matrix.
In other words, $\Phi$ itself is dependent on the unknown signal $x$.
This dependency is not only unusual in \ac{CS}, it is also contrary to what makes a measurement matrix favorable in \ac{CS}: while random matrices tend to capture a representative sample of signal entries, both large and small, the clipping process deliberately excludes all of the large signal entries and keeps only the small ones.

\section{Trivial Pursuit with Clipping Constraints}\label{sec:tpcc}

\begin{figure}[tb]
  \centering
  \subfloat[\label{fig:suppEstimationa}]{
  \includegraphics[width=0.5\columnwidth,page=1]{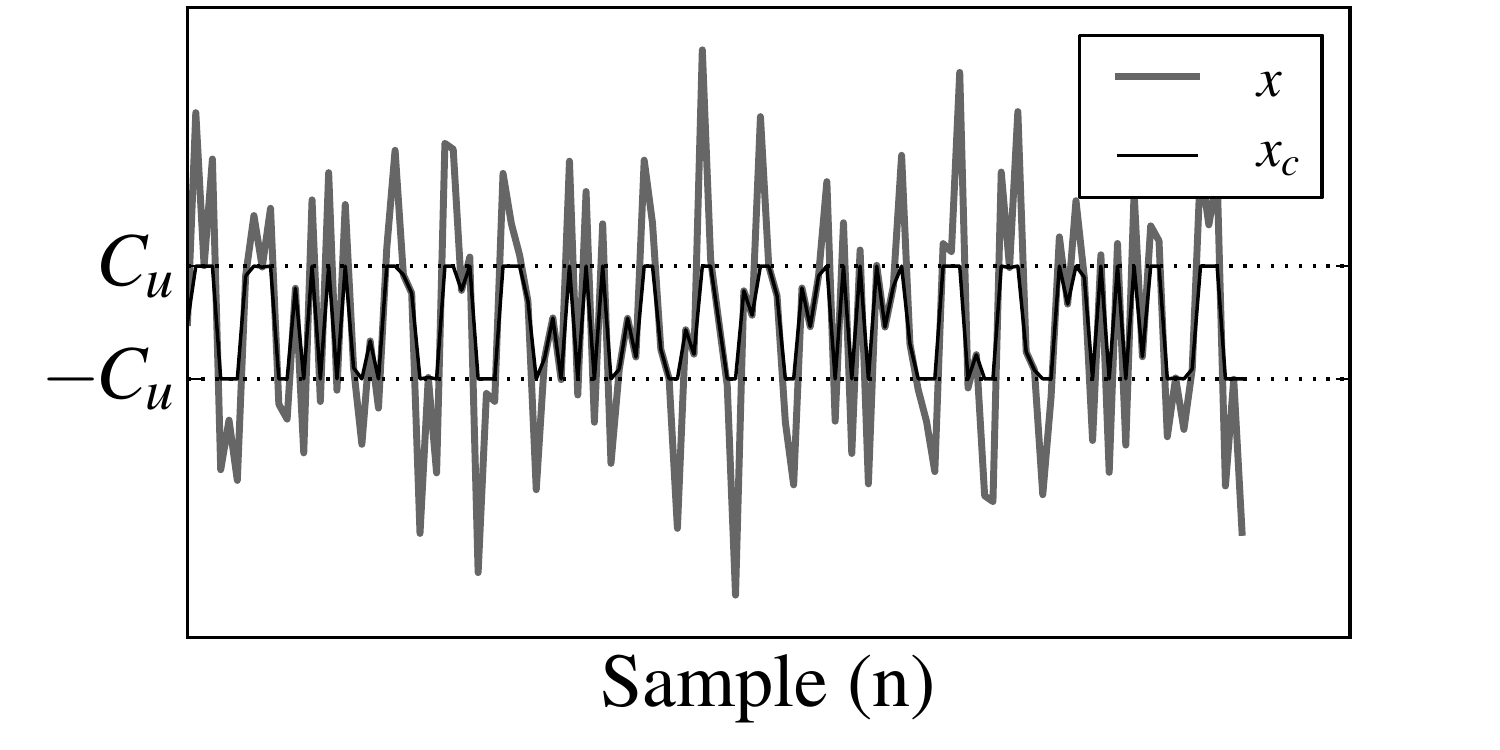}}
  \subfloat[\label{fig:suppEstimationb}]{
  \includegraphics[width=0.5\columnwidth,page=2]{supp_estimation}}
\caption{\small\sl Support estimation using the \ac{DFT} of the clipped signal. (a)~A signal $x$ with sparsity level $K=10$ and its clipped version $x_c$, with $C_u/\|x\|_\infty=0.2$, corresponding to $M=40$ non-clipped samples. (b)~\ac{DFT} of $x$ and $x_c$ for $0\leq k <\frac{N}{2}$. Note that the 5 biggest harmonics of $x_c$ are at the same locations as the harmonics of $x$.}
  \label{fig:suppEstimation}
\end{figure}

Let us note that the \ac{DFT} of the clipped signal $x_c$ contains, in addition to the harmonics introduced by the clipping process, all of the harmonics present in the original signal $x$.
Interestingly, the harmonics with the biggest magnitude typically coincide with the ones from the original signal.
Figure~\ref{fig:suppEstimation} shows an example for a signal with sparsity level $K=10$ clipped at an amplitude corresponding to 20\% of its peak value.
We see that the 5 biggest harmonics of $x_c$ are at the same locations as the 5 harmonics of $x$.
Our second proposed de-clipping algorithm exploits this observation.

The method is very simple and consists of two stages. First, we identify the support (the location of the non-zero Fourier coefficients) of the signal. %
Second, we estimate the value of the coefficients on this support using a least-squares approach, similar to that used in other greedy methods such as Matching Pursuit or \ac{OMP}~\cite{Elad2010a}.
If we know the sparsity level $K$ {\em a priori}, we can estimate the support simply by finding the $K$ biggest harmonics of $x_c$.
In the more general case where we do not know $K$, we select the elements of the support one at a time in a greedy manner, until the reconstruction error on the non-clipped samples is small enough.

Algorithm \ref{alg:TPCC} shows a detailed description of the method.
In the \textbf{match} step we compute the \ac{DFT} of the clipped signal---{\em this happens only once}.
Then we repeat the following steps until the residual $r$ is arbitrary small (we use $\epsilon=10^{-6}$ in our experiments).
In the \textbf{identify} step we add the indices associated with the current largest harmonic to the support index set $\Lambda$, and we then set those coefficients to zero to avoid selecting them again in the next iteration.
In the \textbf{update} step we compute the \ac{DFT} coefficients of a signal---restricted to the support $\Lambda$---that best approximates the non-clipped samples $y$ in a least-squares sense.
Note that the coefficients $\alpha_\Lambda$ on this support are easily computed as $\alpha_\Lambda = (\Phi\Psi)_\Lambda^\dagger y$, where $(\Phi\Psi)_\Lambda^\dagger$ is the pseudoinverse of the columns of $\Phi\Psi$ indexed by $\Lambda$.

\begin{algorithm}[t]
\caption{Trivial Pursuit with Clipping Constraints} \label{alg:TPCC}
\fontsize{9}{9}\selectfont 
\begin{algorithmic}
\REQUIRE  $\Phi$, $\Psi$, $x_c$. $y$, $\epsilon$
\STATE \textbf{initialize:} $r = y$, $\Lambda^{(1)} =
  \emptyset$, $\ell = 1$
\STATE \textbf{match:} $h = \DFT\{x_c\}$
       (indexed from 0 to $N-1$)
\WHILE{$\|r\|_2 > \epsilon$}
\STATE
\begin{tabular}{ll}
  \textbf{identify:} & $k = \argmax_{0\leq j\leq\frac{N}{2}} |h(j)|$ \\
  & $\Lambda^{(\ell+1)} = \Lambda^{(\ell)} \cup \{k, (N-k) \mod N\}$ \\
  & $h(k) = 0 $\\
  \textbf{update:} & $\alpha = \argmin_{z: ~ \supp(z) \subseteq
    \Lambda^{(\ell+1)}}  \|y - \Phi \Psi z\|_2$ \\
  & $r = y - \Phi \Psi \alpha$   \\
  & $\ell = \ell+1$
\end{tabular}
\ENDWHILE
\ENSURE $\widehat{x} = \Psi\alpha = \Psi\argmin_{z: ~ \supp(z) \subseteq
  \Lambda^{(\ell)}} \|y - \Phi \Psi z\|_2$
\end{algorithmic}
\end{algorithm}

\ifthenelse {\boolean{Draft}}
{
\begin{figure}[tb]
\begin{minipage}[b]{0.5\linewidth}
\centering
  \includegraphics[width=\columnwidth]{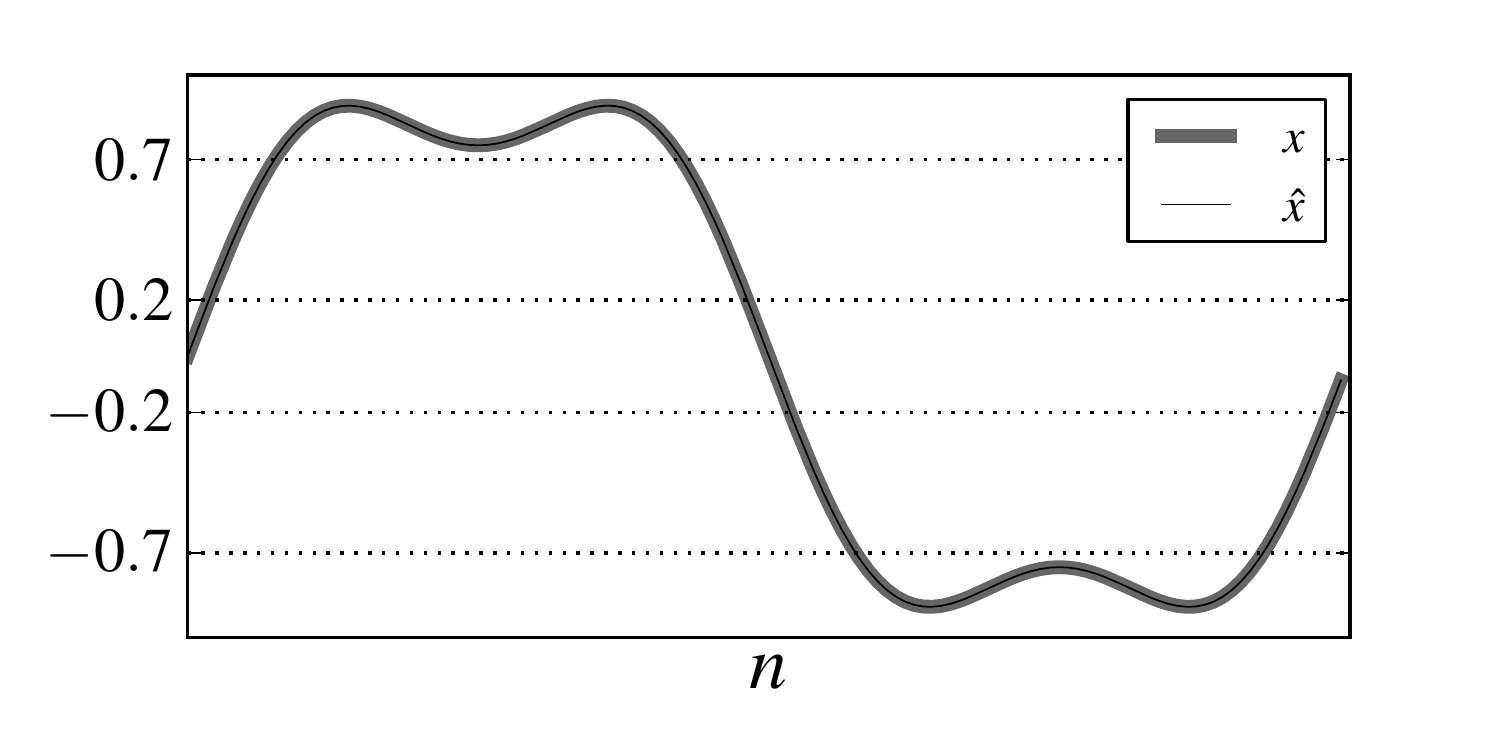}
  \vspace{0.0ex}
  \caption{\small\sl Recovering a two-tone signal using \ac{TPCC}. The
    clipping level $C_u=0.7$ is just below the high-frequency ``bumps''. It is
    possible to recover the signal with a clipping level down to $C_u=0.2$. We
    set the signal length to $N=128$.}
  \vspace{0.0ex}
  \label{fig:TPCC}
\end{minipage}
\hspace{0.5cm}
\begin{minipage}[b]{0.5\linewidth}
\centering

\includegraphics[scale=0.4]{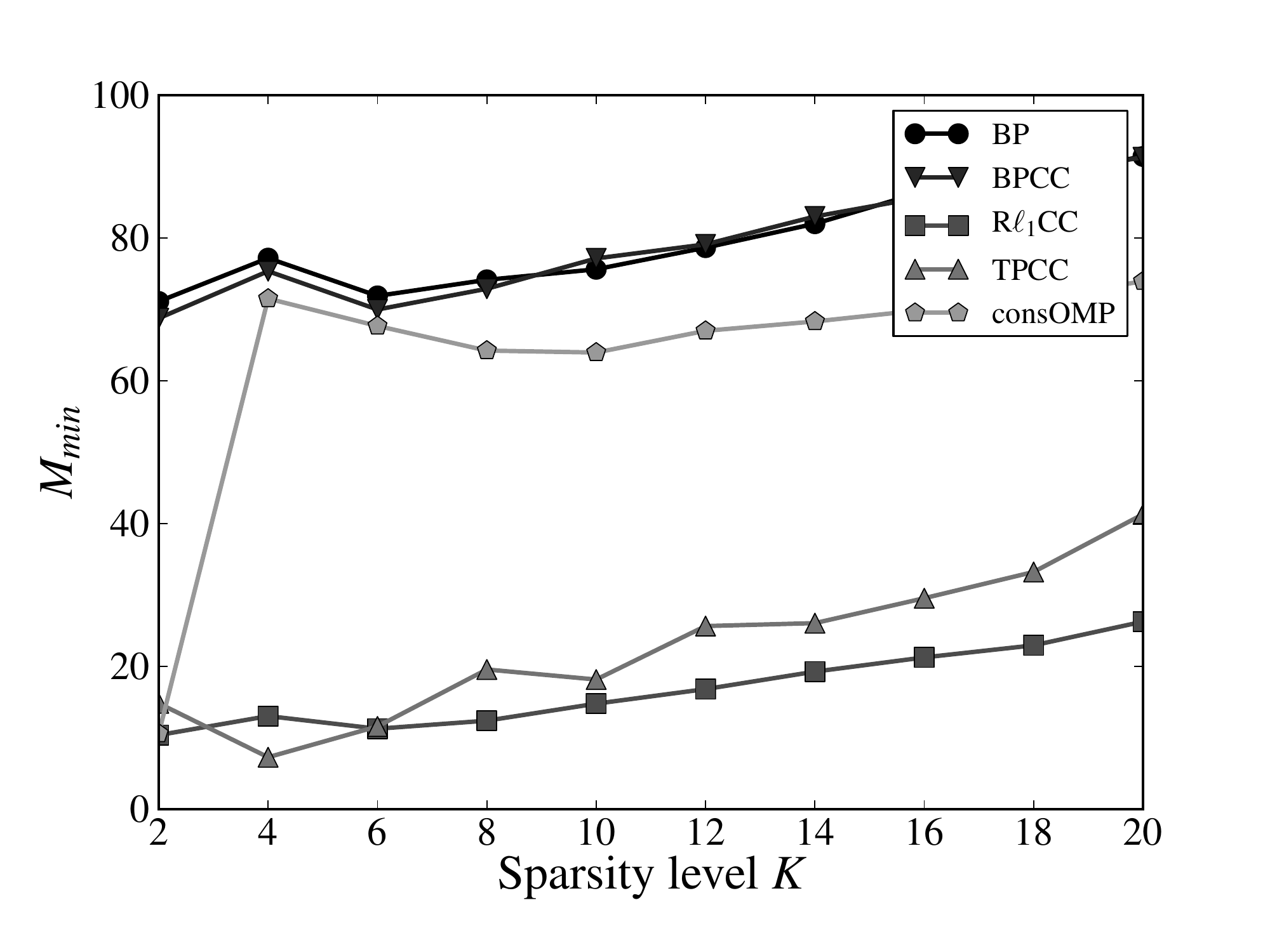}

\caption{\small\sl Recovering a clipped signal using BP, BPCC,
  \emph{constraint}-\ac{OMP}, \ac{Rell1}, and \ac{TPCC}. We plot the average
  minimum number of non-clipped samples $M_{min}$ required to recover signals
  of different sparsity levels $K$.}
\vspace{3.8ex}
\label{fig:exp1TPCC}
\end{minipage}
\end{figure}
}{}

Although perhaps not evident at first sight, Algorithm \ref{alg:TPCC} corresponds to a modified version of the method known as Trivial Pursuit (TP)~\cite{Elad2010a,Baron2005}.
Given a set of non-clipped linear measurements $y = \Phi x = \Phi \Psi \alpha$, TP would estimate the support of $\alpha$ simply by computing the score $h_{TP} = (\Phi\Psi)^T y$
and selecting the indices of the largest entries of $h_{TP}$.
We can write $h_{TP} = (\Phi\Psi)^T y = \Psi^T\Phi^Ty$ and note that the vector $\Phi^T y \in \real^N$ corresponds to a zero-padded version of $y$ with the non-clipped samples at the proper locations.
Since multiplying by $\Psi^T$ is equivalent to computing the \ac{DFT} of a vector, $h_{TP}$ is in fact the \ac{DFT} of the zero-padded version of $y$.
The vector $h$ computed in the \textbf{match} of Algorithm \ref{alg:TPCC} is actually very similar to $h_{TP}$, except that instead of computing the \ac{DFT} of the zero-padded version of $y$, we compute the \ac{DFT} of $x_c$, which is equal to $y$ padded with the clipped values instead of zeros.
In other words, Algorithm \ref{alg:TPCC} exploits the knowledge of the clipped values.
For this reason we dub our method \ac{TPCC}.

To illustrate the effectiveness of TPCC we experiment with the signal $x(n) =  \sin(2\pi n/N) + 0.25\sin(2\pi 3n/N)$
of length $N=128$.
We clip this signal, shown in Fig. \ref{fig:TPCC}, at a level just below the ``bumps''.
It might seem impossible to recover the signal once the oscillations due to the third harmonic are missing.
Remarkably, however, \ac{TPCC} not only recovers this signal at the clipping level of $C_u=0.7$, but it can even recover this signal down to the clipping level of $C_u=0.2$, at which point there are only 10 non-clipped samples.

\ifthenelse {\boolean{Draft}}{}{
\begin{figure}[tb]
  \centering
  \includegraphics[width=0.65\columnwidth]{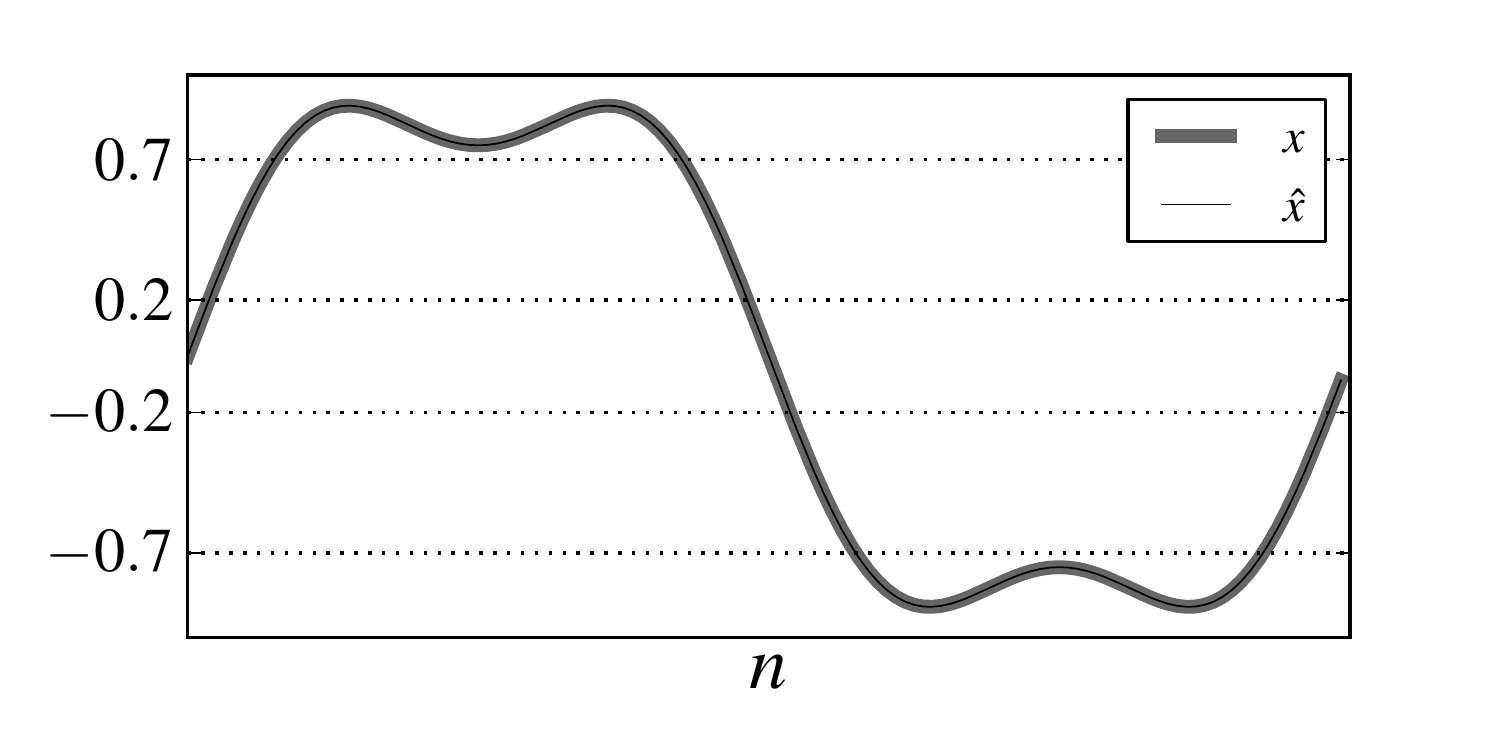}
  \caption{\small\sl Recovering a two-tone signal using \ac{TPCC}. The
    clipping level $C_u=0.7$ is just below the high-frequency ``bumps''. It is
    possible to recover the signal with a clipping level down to $C_u=0.2$. We
    set the signal length to $N=128$. \vspace{-0.2in}}
  \label{fig:TPCC}
\end{figure}
}

Although \ac{TP} is arguably the simplest reconstruction method for sparse signals in \ac{CS}, it also is one of the poorest performing in terms of the number of measurements required for successful signal recovery~\cite{Elad2010a}.
For this reason it is quite surprising that in this experiment and in others (see Sec.~\ref{sec:experiments}) \ac{TPCC} can be so effective for de-clipping sparse signals.

\section{Experimental Results \label{sec:experiments}}

In this section we empirically evaluate the methods described previously. We also compare with \emph{constraint}-\ac{OMP}.\footnote{We have found that
\emph{constraint}-\ac{OMP} exhibits better performance with signals sparse in
the \ac{DCT} domain than with signals sparse in the \ac{DFT} domain. We thus
use the \ac{DCT} as the sparsity basis for testing this method.} For all
experiments that follow we generate, for each value of the sparsity level $K$,
signals of length $N=128$ having $K$ non-zero coefficients with frequencies
selected randomly, amplitudes chosen randomly from a uniform distribution
between 0.5 and 1.5, and phases selected randomly.\footnote{\textsc{Matlab} code is available at https://github.com/aweinstein/declipping.}

In the first experiment, we find the average minimum number $M_{min}$ of non-clipped samples required to recover a signal as a function of $K$. We compute the average over 100 simulation runs. Figure~\ref{fig:exp1TPCC} shows the results. BP and BPCC perform very poorly, being unable to recover the original signal except when the clipping is very mild. \emph{Constraint}-\ac{OMP} performs better, while \ac{TPCC} and \ac{Rell1} perform much better still. In fact, both \ac{TPCC} and \ac{Rell1} can reliably recover the signal using a number of non-clipped samples that is not much larger than $K$, while BP and BPCC require a number of non-clipped samples much closer to $N$.

\newcommand{\figwidth}{0.78}

\ifthenelse {\boolean{Draft}}{}{
\begin{figure}[tb]
  \centering
  \includegraphics[width=\figwidth\columnwidth]{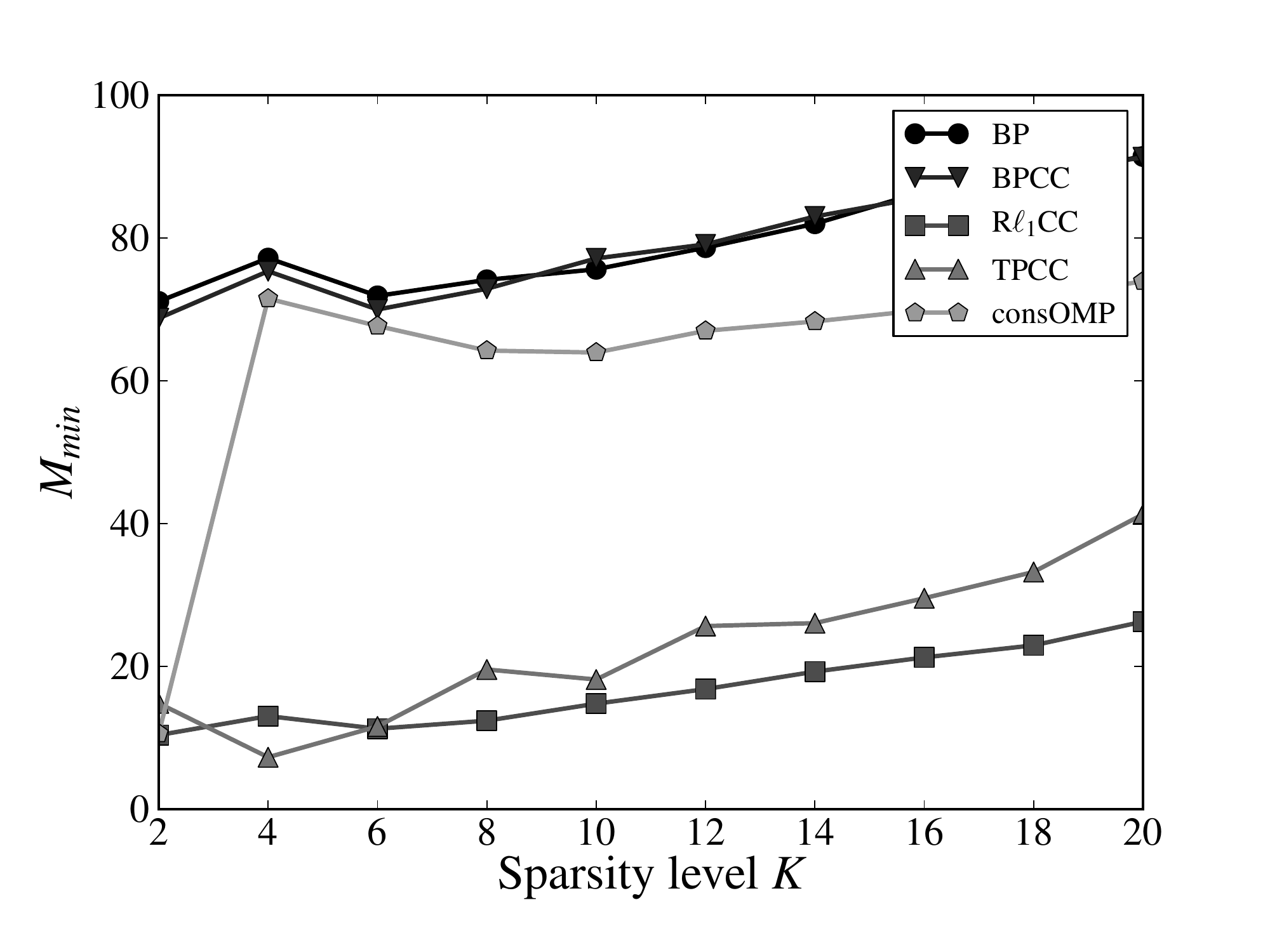}
  \caption{\small\sl Recovering a clipped signal using BP, BPCC, \emph{constraint}-\ac{OMP}, \ac{Rell1}, and \ac{TPCC}. We plot the average minimum number of non-clipped samples $M_{min}$ required to recover signals of different sparsity levels $K$. \vspace{-0.2in}}
  \label{fig:exp1TPCC}
\end{figure}
}

\ifthenelse {\boolean{Draft}}{
\begin{figure}[tb]
\begin{minipage}[b]{0.5\linewidth}
\centering
\includegraphics[scale=0.4]{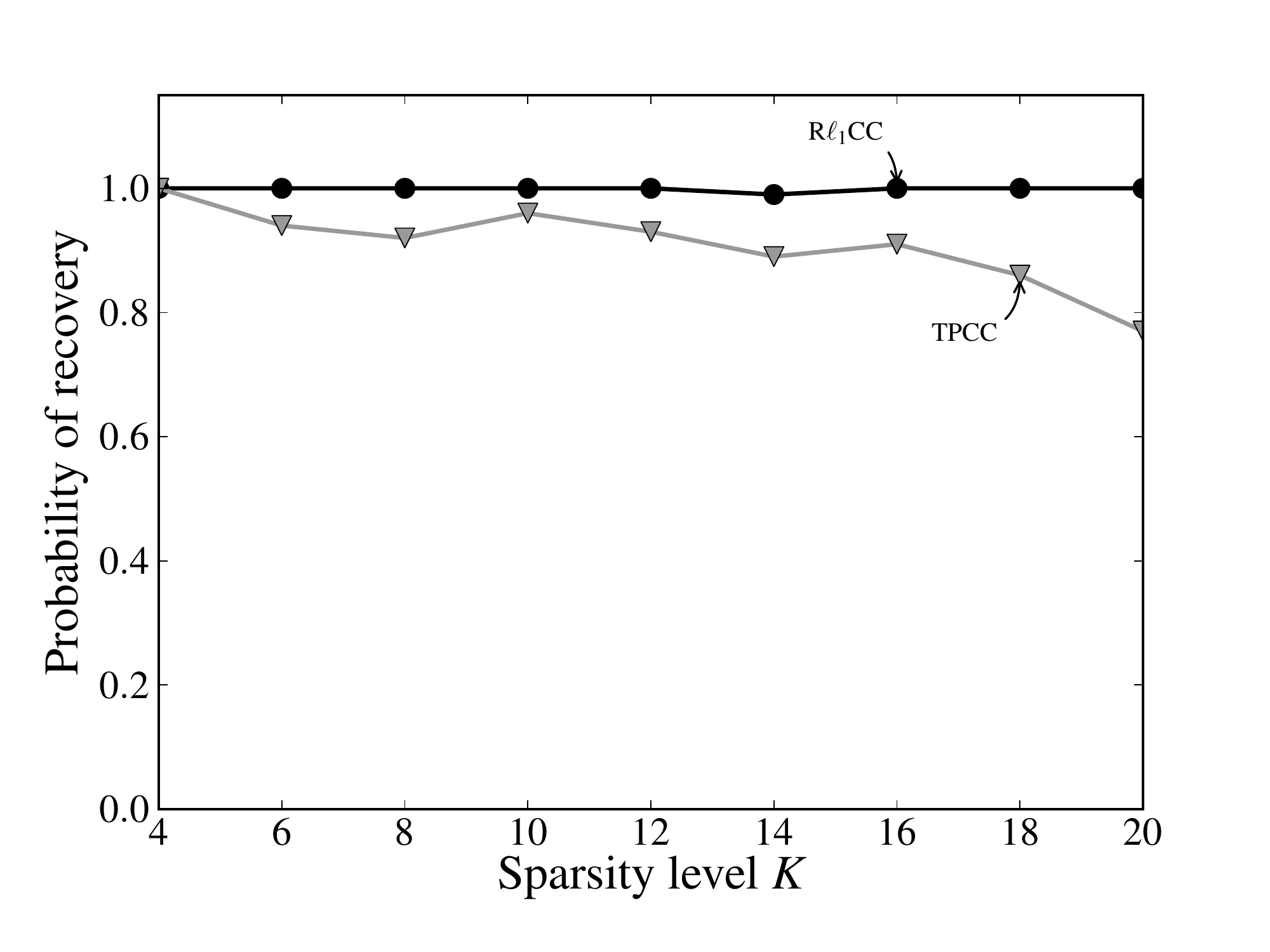}
\caption{\small\sl Recovering a clipped signal using \ac{Rell1} and \ac{TPCC}. We plot the probability of perfect recovery as a function of the sparsity level $K$ for $M=70$ non-clipped samples.}
\label{fig:exp2TPCC}
\end{minipage}
\hspace{0.5cm}
\begin{minipage}[b]{0.5\linewidth}
\centering
\includegraphics[scale=0.4]{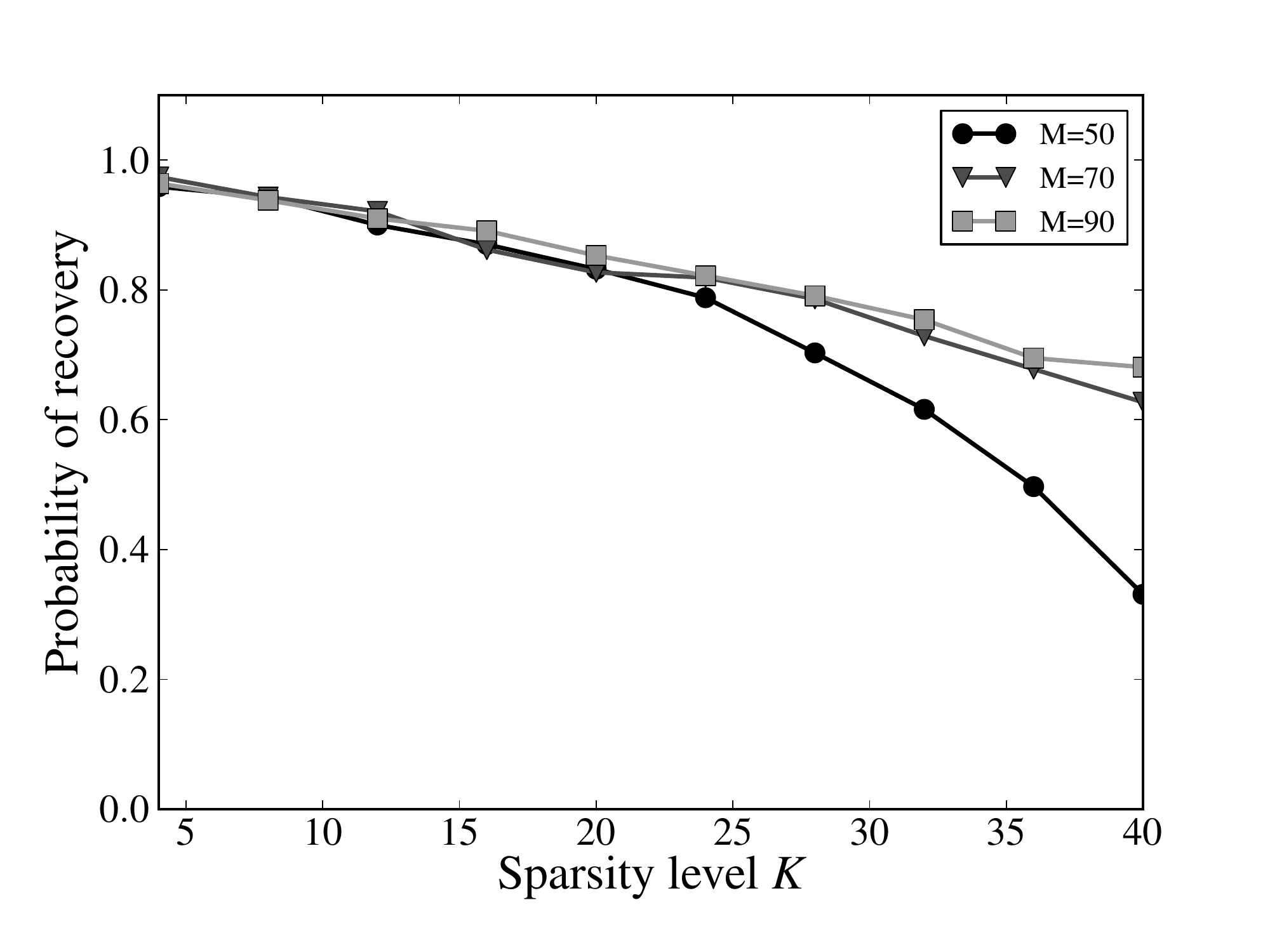}
\caption{\small\sl Recovering a clipped signal using \ac{TPCC}. We plot the probability of perfect recovery as a function of the sparsity level $K$ for different numbers of non-clipped samples $M$.}
\label{fig:exp3TPCC}
\end{minipage}
\end{figure}
}{}

In the second experiment, we compare \ac{Rell1} and \ac{TPCC} in a different way. We fix the number of non-clipped samples to $M=70$ and plot the
probability of perfect recovery (declared when $\|x-\widehat{x}\|\leq 10^{-3}$) as a function of the sparsity level $K$. Figure~\ref{fig:exp2TPCC} shows the results using 100 trials for each combination of $M$ and $K$. Although \ac{Rell1} performs somewhat better than \ac{TPCC}, it is important to underscore that \ac{TPCC} requires significantly fewer computations.

\ifthenelse {\boolean{Draft}}{}{
\begin{figure}[tb]
  \centering
  \includegraphics[width=\figwidth\columnwidth]{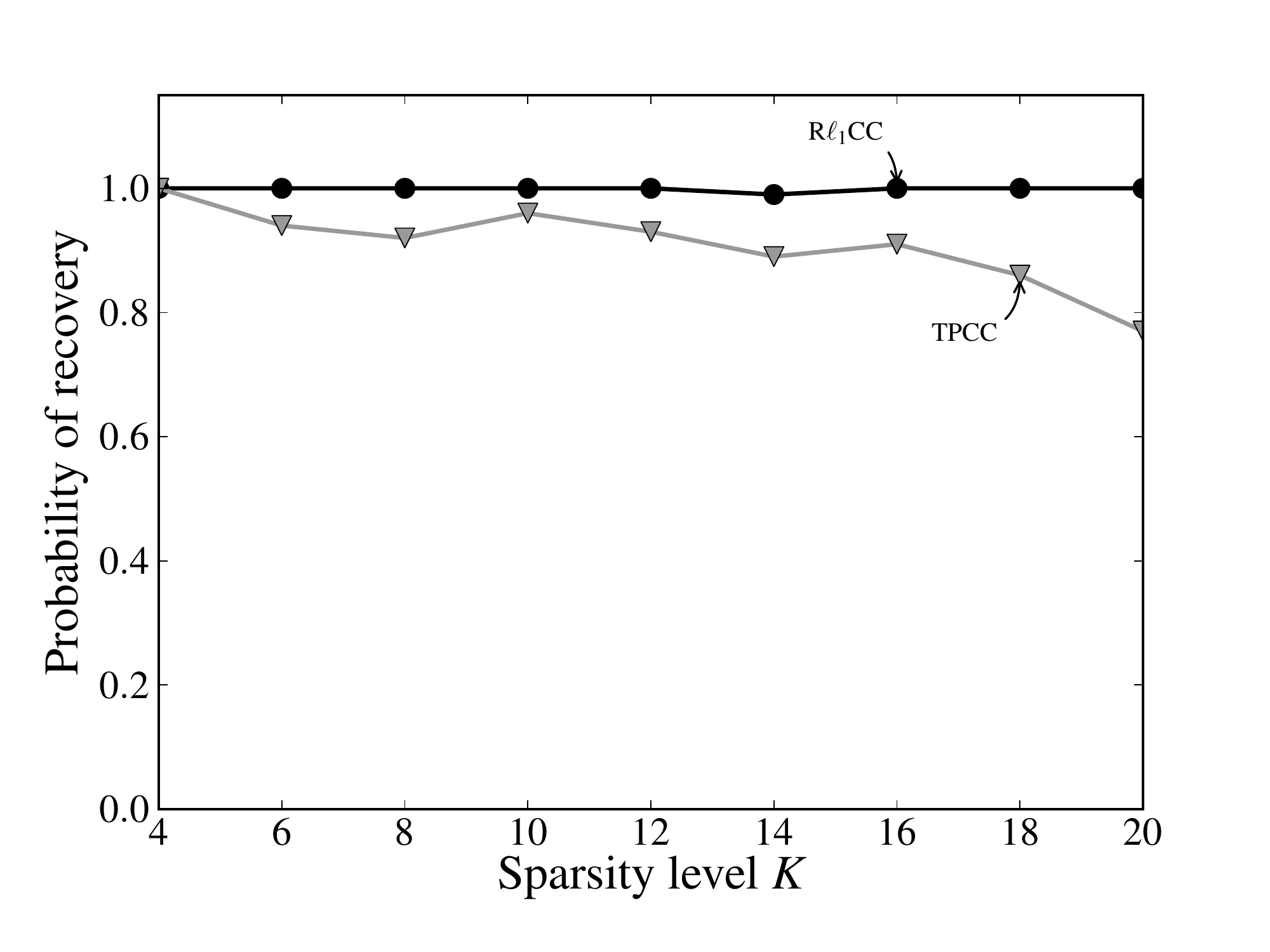}
  \caption{\small\sl Recovering a clipped signal using \ac{Rell1} and \ac{TPCC}. We plot the probability of perfect recovery as a function of the sparsity level $K$ for $M=70$ non-clipped samples. \vspace{-0.2in}}
  \label{fig:exp2TPCC}
\end{figure}
}

In the final experiment, we examine the performance of \ac{TPCC} more closely. We plot the probability of perfect recovery as a function of $K$ for different values of $M$. Figure~\ref{fig:exp3TPCC} shows the results using 500 simulation runs for each combination of $M$ and $K$. As expected, we the probability of recovery increases as the sparsity level $K$ decreases, and as the number of non-clipped samples $M$ increases. Again, in general, we can expect a high probability of recovery from \ac{TPCC} (over our random signal model) when $M$ is a small multiple of $K$.

\ifthenelse {\boolean{Draft}}{}{
\begin{figure}[tb]
  \centering
  \includegraphics[width=\figwidth\columnwidth]{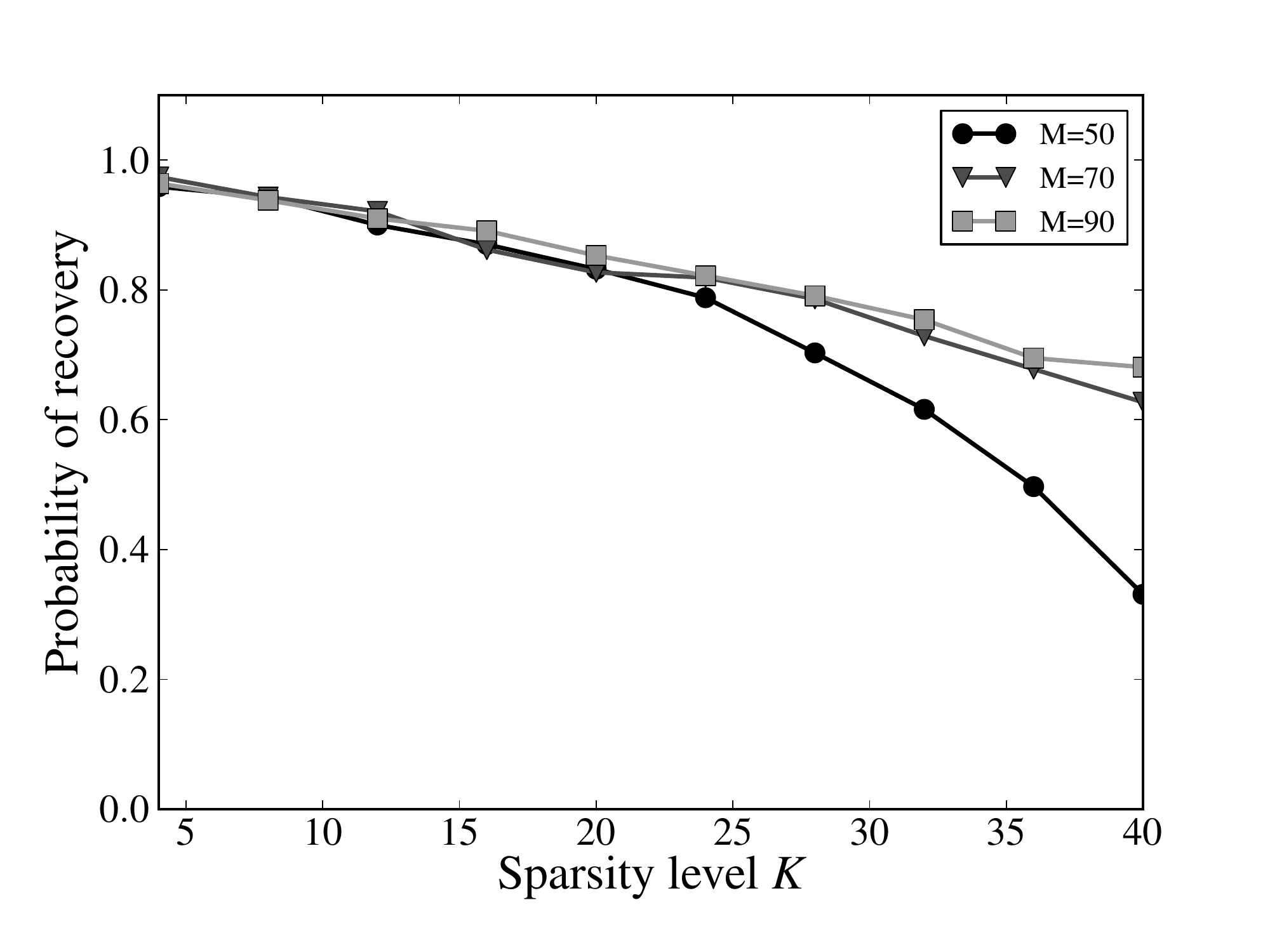}
  \caption{\small\sl Recovering a clipped signal using \ac{TPCC}. We plot the probability of perfect recovery as a function of the sparsity level $K$ for different numbers of non-clipped samples $M$. \vspace{-0.2in}}
  \label{fig:exp3TPCC}
\end{figure}
}

\newboolean{Noisy}
\setboolean{Noisy}{false}

\ifthenelse{\boolean{Noisy}}{
It is also possible to extend \ac{TPCC} to the case where the original signal
$x$ is contaminated by additive noise. All we need to do, as commonly done in
greedy methods, is to modify the value of $\epsilon$ used in the stopping
condition of Algorithm \ref{alg:TPCC} as a function of the noise level. We
consider the noisy signal $x_n = x + z$, where $z$ is a bounded noise term with
$\|z\|_2<\delta$. We observe the clipped signal $x_c$ equal to $x_n$ if $C_l <
x_n < C_u$, equal to $C_u$ if $x_n>C_u$, and equal to $C_l$ if $x_n<C_l$.

We evaluate the effectiveness of this approach with two signal
realizations. Both signals have sparsity level $K$ equal to 10 and fixed support
(equal to the 10 lowest frequencies). The amplitude of the harmonics are drawn
from a uniform distribution between 0.5 and 1.5. We fix $\|z\|_2$ equal to 1
and $\epsilon$ equal 1. Figure \ref{fig:TPCC_noisy}\subref{fig:tpcca} shows the
simulation result for a signal with 54 non-clipped samples, and Fig.
\ref{fig:TPCC_noisy}\subref{fig:tpccb} shows the result for a signal with 46
non-clipped samples.

\begin{figure}[tb]
  \centering
  \subfloat[\label{fig:tpcca}]{
    \includegraphics[width=0.5\columnwidth]{tpcc_noisy_a}
  }
  \subfloat[\label{fig:tpccb}]{
    \includegraphics[width=0.5\columnwidth]{tpcc_noisy_b}}
  \caption{\small\sl Recovering a noisy clipped signal using \ac{TPCC}. We plot the
    original noisy signal $x+z$ and the recovered signal $\widehat{x}$ for two
    signal realizations. We fix the signal length to $N=128$, the noise norm to
    $\|z\|_2=1$ and the stopping criteria to $\epsilon=1$. The number of
    non-clipped samples is \subref{fig:tpcca} 54 and \subref{fig:tpccb}
    46. \label{fig:TPCC_noisy} \vspace{-0.2in}}
\end{figure}
}{}

\section{Conclusions}\label{sec:conclusions}

We have presented two methods for restoring a clipped signal using the model assumption of sparsity in the frequency domain.
Both techniques can be extended to account for noise, and preliminary experiments on noisy signals are promising although space limitations prevent us from discussing this more deeply.
One of our methods, \ac{TPCC}, is particularly simple to implement; its running time is dominated by the computation of the \ac{DFT} and
the solution of the least-squares problem, and it is significantly faster than \ac{Rell1}.

Our algorithms are inspired by existing techniques from the field of \ac{CS}, and the performance we achieve (where the requisite number of non-clipped samples $M$ scales with $K$) is fully in line with the state-of-the-art performance in \ac{CS}.
This is in spite of the fact that standard RIP analysis does not apply to the de-clipping problem and that the measurement operator in our problem is non-random and signal-dependent.
Insight from \ac{CS} would suggest that this signal dependence could be catastrophic for standard sparse approximation algorithms.
Thus, we believe that further work is warranted to understand why even a simple algorithm such as \ac{TPCC} can succeed in the de-clipping problem when much more complicated algorithms are required in \ac{CS}.

\bibliographystyle{IEEEtran}
\bibliography{library}

\begin{thebibliography}{1}
\providecommand{\url}[1]{#1}
\csname url@samestyle\endcsname
\providecommand{\newblock}{\relax}
\providecommand{\bibinfo}[2]{#2}
\providecommand{\BIBentrySTDinterwordspacing}{\spaceskip=0pt\relax}
\providecommand{\BIBentryALTinterwordstretchfactor}{4}
\providecommand{\BIBentryALTinterwordspacing}{\spaceskip=\fontdimen2\font plus
\BIBentryALTinterwordstretchfactor\fontdimen3\font minus
  \fontdimen4\font\relax}
\providecommand{\BIBforeignlanguage}[2]{{%
\expandafter\ifx\csname l@#1\endcsname\relax
\typeout{** WARNING: IEEEtran.bst: No hyphenation pattern has been}%
\typeout{** loaded for the language `#1'. Using the pattern for}%
\typeout{** the default language instead.}%
\else
\language=\csname l@#1\endcsname
\fi
#2}}
\providecommand{\BIBdecl}{\relax}
\BIBdecl

\bibitem{Abel1991}
J.~S. Abel and J.~O. Smith, ``{Restoring a clipped signal},'' in \emph{Proc.
  Int. Conf. Acoustics, Speech and Signal Processing (ICASSP)}, 1991.

\bibitem{Godsill2001}
S.~J. Godsill, P.~J. Wolfe, and W.~N. Fong, ``{Statistical model-based
  approaches to audio restoration and analysis},'' \emph{J. New Music Res.},
  vol.~30, no.~4, pp. 323--338, Dec. 2001.

\bibitem{Olofsson2005}
T.~Olofsson, ``{Deconvolution and model-based restoration of clipped ultrasonic
  signals},'' \emph{IEEE Trans. Instrum. Meas.}, vol.~54, no.~3, pp.
  1235--1240, Jun. 2005.

\bibitem{Candes2006a}
E.~J. Cand\`{e}s, ``{Compressive sampling},'' in \emph{Proc. Int. Congr.
  Math.}, vol.~17, no.~4, Spain, Jan. 2006.

\bibitem{Elad2010a}
M.~Elad, \emph{{Sparse and Redundant Representations}}.\hskip 1em plus 0.5em
  minus 0.4em\relax New York, NY: Springer, 2010.

\bibitem{Gemmeke2010}
J.~F. Gemmeke, H.~{Van Hamme}, B.~B. Cranen, and L.~Boves, ``{Compressive
  Sensing for missing data imputation in noise robust speech recognition},''
  \emph{IEEE J. Sel. Top. Signal Process.}, vol.~4, no.~2, pp. 272--287, Apr.
  2010.

\bibitem{Adler2011}
A.~Adler, V.~Emiya, M.~G. Jafari, M.~Elad, R.~Gribonval, and M.~D. Plumbley,
  ``{A Constrained Matching Pursuit approach to audio declipping},'' in
  \emph{Proc. Int. Conf. Acoustics, Speech and Signal Processing (ICASSP)},
  2011.

\bibitem{Candes2008b}
E.~J. Cand\`{e}s, M.~B. Wakin, and S.~P. Boyd, ``{Enhancing sparsity by
  reweighted $\ell_1$ minimization},'' \emph{J. Fourier Anal. Appl.}, vol.~14,
  no.~5, pp. 877--905, Nov. 2008.

\bibitem{Baron2005}
\BIBentryALTinterwordspacing
D.~Baron, M.~F. Duarte, M.~B. Wakin, S.~Sarvotham, and R.~G. Baraniuk,
  ``{Distributed Compressive sensing},'' \emph{Preprint}, Jan. 2009. [Online].
  Available: \url{http://arxiv.org/abs/0901.3403}
\BIBentrySTDinterwordspacing

\end{thebibliography}

\end{document}